# Pseudo Quantum Random Number Generator with Quantum Permutation Pad

Randy Kuang
Quantropi Inc.
Ottawa, Canada
randy.kuang@quantropi.com

Dafu Lou
Quantropi Inc.
Ottawa, Canada
dafu.lou@quantropi.com

Alex He
Quantropi Inc.
Ottawa, Canada
alex.he@quantropi.com

Chris McKenzie
Quantropi Inc.
Ottawa, Canada
chris.mckenzie@quantropi.com

Michael Redding
Quantropi Inc.
Ottawa, Canada
michael.redding@quantropi.com

*Abstract*— **Cryptographic random number generation is critical for any quantum safe encryption. Based on the natural uncertainty of some quantum processes, variety of quantum random number generators or QRNGs have been created with physical quantum processes. They generally generate random numbers with good unpredictable randomness. Of course, physical QRNGs are costic and require physical integrations with computing systems. This paper proposes a pseudo quantum random number generator with a quantum algorithm called quantum permutation pad or QPP, leveraging the high entropy of quantum permutation space its bijective transformation. Unlike the Boolean algebra where the size of information space is $2^n$ for an n-bit system, an n-bit quantum permutation space consists of $2^n!$ quantum permutation matrices, representing all quantum permutation gates over an n-bit computational basis. This permutation space holds an equivalent Shannon information entropy $\log_2(2^n!)$. A QPP can be used to create a pseudo QRNG or pQRNG capable integrated with any classical computing system or directly with any application for good quality deterministic random number generation. Using a QPP pad with 64 8-bit permuation matrices, pQRNG holds 107,776 bits of entropy for the pseudo random number generation, comparing with 4096 bits of entropy in Linux /dev/random. It can be used as a deterministic PRNG or entropy booster of other PRNGs. It can also be used as a whitening algorithm for any hardware random number generator including QRNG without discarding physical bias bits.**

I. INTRODUCTION

Random number generations can be categorized into two classes: hardware random number generators or TRNGs and software/pseudo random number generators or PRNGs. A HRNG is a device which generates random numbers from a specific physical process such as noise sampling, free running oscillators, chaos, and quantum effects. These processes are generally considered to be unpredictable. Of HRNGs, quantum random number generators are specifically referred to optical devices. Rarity, Owens, and Tapster (1994) [1] reviewed the early status of interferometry-based quantum cryptography and compared photon-pair and faint-pulse schemes. Stefanov, et al (2000) [2] reported their optical quantum random number generator, a simple beam splitter. The random events are realized from the choice of single photons between two outputs of a beam splitter. Ma, et al (2016) published their recent review of quantum random number generators [3]. They classified QRNGs into three categories: practical QRNG, self-testing QRNG, and semi-self-testing QRNG. The practical QRNG is built on fully trusted and calibrated devices and produce good randomness at high speed. The self-testing QRNG generates verifiable randomness without trusting the actual implementation. The semi-self-testing QRNG provides a tradeoff between the trustworthiness on the device and the generation speed. Gehring, et al in 2020, reported their ultra fast quantum random number generation at a speed 8 Gbps based on quadrature measurements of vacuum fluctuations [4]. Gehring, et al in 2021, reported their homedyne-based quantum random number generator at 2.9 Gbps. By using a different technique for quantum random number generation through measurements of laser phase fluctuations. Nie, et al in 2015 [6], reported their extremely high generation speed at 68 Gbps.

Some commercially available QRNGs can be found in the market. ID Quantique's Quantis QRNG offers two form factors of PCI card and USB [7], coming with generation speed at 4 Mbps and 16 Mbps. Quintessence Labs offers their QRNG qStream PCIe card with 8 Gbps quantum entropy source, reduced to 1 Gbps unconditional entropy after the whitening algorithm. Commercial QRNGs usually comes with certint whitening algorithm to remove biases in the outputs of a physical generator.

Although QRNGs can produce truly unpredictable random numbers, they are generally expensive and also not suitable to integrate in certain computing systems such as user end devices. The most common way to have a good randomness generator is to use pseudo-random number generators. James and Moneta (2020) [9] reviewed pseudo-random number generators based on the Kolmogorov–Anosov theory of mixing in classical mechanical systems. Orúe, et al (2017) [10], reported their deep review on cryptographic secure PRNGs for IoT devices. In 2019, Baldanzi, et al, presented a cryptographically secure PRNG based on SHA2 hash algorithm [11]. They analysised different cryptographic algorithms such as SHA2, AES-256 CTR mode, and triple DES to build deterministic random bit generators or DRBGs.

The highest security strength is 256 bits of entropy. Their DRBG based on SHA256 cryptographic primitive has passed NIST randomness testing with high pass rate. They has implemented it on FPGA and ASIC standard-cell technologies. With those hardware accelerations, their cryptographic secure PRNGs demonstrate high throughput pseudo-random number generations. Mandal, et al (2013) [12] designed and analysised a new lightweight cryptographic pseudo random number generator called Warbler PRNG for smart devices, which demonstrates a good randomness and passes NIST randomness testing suite. However, it only has a security of 45-bit entropy.

Among all existing PRNGs, xorShift worthes a special mention although it is generally among the non-cryptographically secure random number generator. Marsaglia created it in 2003 [13], there have been developed multiple variations of improvements such as xorshift* to use an invertible multiplication to its outputs, xorshift+ (64+ or 128+) to use addition for faster non-linear transformations, xoshiro, and xoroshiro with rotations in addition to additions. The unique benefit from xorshift family PRNGs is their fast generation speed. They can simply generate pseudo random numbers at a speed of Giga bytes per second. Vigna (2016) [15] analysised xorshift PRNGs and found xorshift128+ to be the fastest generator successfully passing BigCrush testing.

One of major issues from existing PRNGs is the limited entropy injected with a seed. With our knowledge, the highest entropy accepted by a PRNG algorithm is 1024 bits in xorshift1024+/xorshift1024* where statistical tests are also failed for linearity. That indicates that increasing the seed length may not fix those failures.

Kuang and Bettenburg in 2020 [16] proposed a new algorithm based on quantum permutation logic gates or quantum permutation pad or QPP over quantum computational basis. AbdAllah1, Kuang, and Huang also applied QPP to generate Just-in-Time shared keys (JIT-SK) for TLS 1.3 Zero roundtrip time (0-RTT) [17]. Kuang, et al in 2021 [18], proposed a quantum safe lightweigh cryptographic algorithm by replacing SubBytes and AddRoundKey with the same QPP in AES algorithm and achieved a round reduction by two-thirds. Kuang and Barbeau (2021) [19] proposes a universal quantum safe cryptography with QPP. This paper plans to build a pseudo quantum random number generator or pQRNG with QPP, based on a quantum computing algorithm.

In the remaining parts, we will briefly introduce QPP in section 2, then propose pQRNG and perform randomness analysis in section 3, and a conclusion will be drawn at the end.

## II. QUANTUM PERMUTATION PAD

Classical computing systems are built on the Boolean algebra with a set of basic Boolean logic gates such as AND, OR, NAND, NOR, and XOR. They are bitwise operations.

Quantum computers are built on linear algebra over Hilbert space, or called computational basis in quantum computing, with operations represented by quantum logic gates such as Hadamard gate and permutation gates. The mathemathical expressions of quantum logic gates are all unitary and reversable square matrices over the computational basis. Quantum gates are classified into two categories: non-classical behavior and classical behavior gates. The former represents quantum superpositions and entanglements and the later is deterministic transformation from an input state of the system to an output state, or simply a state permutation. For a n-qubit system with $2^n$ information states represented by Galois field $GF(2^n)$, the entire state permutations form the symmetric group $S_{2^n}$, with total $2^n!$ unique permutations. A generic permutation gate can be physically implemented with an algorithm proposed by Shende et al in 2003 [20] using quantum NOT, CNOT and TOFFOLI gates in a quantum computing system and can be also mathematically expressed with permutation matrix in classical computing systems.

A n-qubit permutation gate can be represented by a $2^n \times 2^n$ permutation matrix $P[2^n, 2^n]$ over a quantum computational basis: $\{|0\rangle, |1\rangle, \ldots, |2^n-1\rangle\}$, with only one element to be 1 on each row and each column and all others to be 0. Each permutation matrix represents a bijective mapping from input information space to output space. There exist $2^n!$ unique bijective mappings between input and output information space over the computational basis (note: only $2^n$ mappings under Boolean algebra). The entire permutation matrices form a special space called permtation space of $2^n!$ dimensions, associated with an equivalent Shannon entropy $e = \log_2(2^n!) \approx 2^n (n - 0.42)$ bits at a larger n. For n=8 bits, the corresponding entropy is 1684 bits. Therefore, an n-bit permutation space can be considered as an entropy expansion from the classical Boolean information space or Galois field $GF(2^n)$ to quantum permutation space or $S_{2^n}$. This huge entropy from the quantum permutation space paves a foundation for quantum safe cryptography with the property of the Shannon perfect secrecy [16].

An n-bit permutation matrix can be randomly selected through the Fisher-Yates shuffling algorithm with a true random seed of length $n2^n$ bits as shown in Algorithm 1 for n = 8. For a QPP pad with M permutation matrices, we can repeat the Algorithm 1 for M times to create the pad with $nM2^n$ bits of random secrets. A typical QPP pad with M=64 and n=8 can have an equivalent Shannon entropy = 107,744 bits. Such a high entropy can be used to build pQRNG.

| Algorithm 1. Pseudo code of QPP mapping from the secret key |
|---|
| *-- only illustrate a single permutation matrix selection*<br>*-- state array S[256] → a permutation matrix P[256][256]*<br>*-- initialize P[256][256] to all zeros*<br>*for i from 0 to 255*<br>    S[i] = i<br>*-- input random key k[N] in bytes with N =256*<br>*for i from 255 down to 1 do*<br>  j = k[i]<br>  swap S[j] and S[i]<br>*for i from 0 to 255*<br>    P[i][S[i]] = 1 |

## III. PSEUDO QUANTUM RANDOM NUMBER GENERATOR

As what we discussed in section II, QPP is a quantum algorithm which can be implemented both in a quantum computing system and a classical computing system. It has been proven to be a quantum-based cryptographic algorithm with the property of Shannon perfect secrecy [16]. It is our motivation to build a new quantum algorithm-based pseudo random number generator or pQRNG.

Figure 1 illustrates a deterministic pQRNG either with a input seed or directly retrieve from the local system such as /dev/random or /dev/urandom in a Linux system. The length of the seed is 64x256 Bytes = 16KB. Therefore, a pQRNG has a theoretical internal states $2^{131,072}$, amazing! The PRNG is seeded with the input seed so it can deterministically produce pseudo random numbers to control a dispatcher to select specific permutation matrix in QPP. The Counter is initialized by the supplied seed too. The feedback from the output and is XORed then randomly dispatched to a certain permutation matrix for transformation. A input byte is dispatched to the permutation matrix with index = x >> 2 or rigth shift 2 bits where x is a pseudo random byte produced by the PRNG. The output from QPP is considered as pseudo quantum random numbers or pQRN.

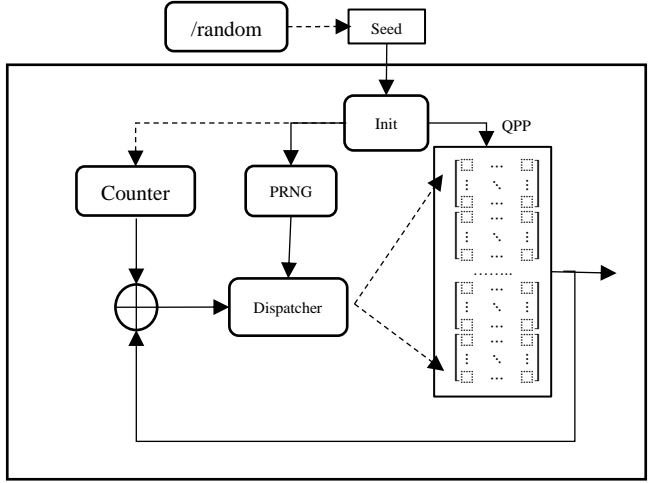

Figure 1. A deterministic pQRNG is illustrated. QPP consists of 64 8-bit permutation matrices to be randomly selected with an external random seed of length up to 16KB.

We use industry recognized randomness testing stuites NIST 800-22, Dieharder, and ENT to test pQRNs from pQRNG. For NIST testing, here are the testing parameters:

- Block frequency: 20,000
- Non-overlapping Template Matching: 9
- Overlapping Template Matching: 9
- Approximate Entropy: 10
- Serial: 10
- Linear Complexity: 500

Table 1 displays the results with NIST 800-22 randomness testing. We generate 1GB random numbers and store it into a binary file and then supply to NIST 800-22 testing suite. For comparisons, we also display testing results together with pseudo random numbers generated from the system rand() through C library and xorshift128+. It is clearly seen from Table 1 that pseudo random numbers generated from both pQRNG and xorshift128+ pass all 15 NIST testing cases. But pseudo random numbers from C library rand() are failed. The same testing results are appeared for Dieharder testing in Table 2. Both xorshift128+ and pQRNG have zero failure but rand() has 2 failures.

Table 1. NIST 800-22 testing reports are illustrated with other two PRNGs. The firt PRNG is from the system standard C library, the second PRNG is xorshift128+ and the third is pQRNG. pQRNG is seeded with 16KB seed. The testing file size is 1GB.

| NIST 800-22 | rand() | xorshift128+ | pQRNG |
|---|---|---|---|
| Frequency | Success | Success | Success |
| Block Frequency | Success | Success | Success |
| Cumulative Sums | Success | Success | Success |
| Runs | Success | Success | Success |
| Longest Run | Success | Success | Success |
| Rank | Success | Success | Success |
| FFT | Failed | Success | Success |
| Non-Overlapping Template | Success | Success | Success |
| Overlapping Template | Failed | Success | Success |
| Universal | Success | Success | Success |
| Approximate Entropy | Success | Success | Success |
| Random Excursions | Success | Success | Success |
| Random Excursions Variant | Success | Success | Success |
| Serial | Success | Success | Success |
| Linear Complexity | Success | Success | Success |

Table 2. Dieharder testing is displayed with the same three PRNGs as in Table 1.

| Dieharder | rand() | xorshift128+ | pQRNG |
|---|---|---|---|
| Passes | 109/114 | 113/114 | 108/114 |
| Weeak | 3/114 | 1/114 | 6/114 |
| Failed | 2/114 | 0 | 0 |

ENT randomness testing suite can generally catch the byte level bias from the supplied random data files. Hurley-Smith, Patsakis and Hernandez-Castro [21] recently identified biased QRNG random generations from a popular commercial QRNG family called Quantis [7] with ENT, where Chi square demonstrates a huge deviation from the idea value 256. In ENT testing, Arithmetical Mean has an ideal value to be 127.50 and Serial Correlation Coefficient measures the extent to which each byte in the file depends upon the previous byte and for true random it should be zero. Monte Carlo $\pi$ indicates the Monte Carlo Value for PI to be ideally 3.14159265. Chi Square should be around 256 with a pvalue between 0.01 and 0.99 for good randomness data. ENT test report with pQRNG is tabulated in Table 3. We illustrate the testing result in Table 3. Again both

xorshift128+ and pQRNG shows very good randomness, especially for Chi square report. The Chi square is 231.04, for pQRNG and 263 for xorshift128+ respectively. However, rand() fails ENT testing with Chi square to be 107.35 and p-value to be 1.0 which indicates that the data is not random for sure although three random generators shows very close testing results for Arithematical Mean, Monte Carlo $\pi$ and Serial Correlation. That is why Chi Square testing can identify if the input data is random or not at byte level.

Table 3. ENT testing is illustrated with the same three PRNGs as in the Table 1.

| ENT | rand() | Xorshift128+ | pQRNG |
|---|---|---|---|
| Entropy (bits) | 8.000000 | 8.000000 | 8.000000 |
| Chi Square | 107.35 | 263.79 | 231.04 |
| p-Value | 1.00 | 0.34 | 0.86 |
| Arith. Mean | 127.5013 | 127.5023 | 127.4995 |
| Monte Carlo $\pi$ | 3.141580069 | 3.141349333 | 3.141659557 |
| Serial Correlation | 0.000052 | 0.000019 | 0.000008 |

One interesting point from Table 3 is the serial correlation value. Of course, the ideal random data should have no correlation to each other. That means, the smaller in the serial correlation is better in randomness. Table 3 shows that the serial correlation is $8 \times 10^{-6}$ from pQRNG, $1.9 \times 10^{-5}$ from xorshift128+ and $5.2 \times 10^{-5}$ from rand(), respectively.

It would be interesting to see the comparison between a physical QRNG and pQRNG. We use a QRNG called qStream from Quintessence Labs. qStream QRNG can generate 1 Giga bits of good random numbers per second, one of the highest throughput on the market. Although both qStream and pQRNG pass NIST and Dieharder randomness test suites, we would like to illustrate the test reports for ENT because ENT randomness test is very sensitive to byte level bias [21]. Table 4 lists three sets of reports, two from pStream and one from pQRNG. For pStream QRNG, we perform ENT randomness testing with 300 MB and 1 GB of random numbers. All three reports pass ENT testing without visible byte level bias. Chi square values are around the ideal value 256, with good p-values. But it is surprisingly noticed that testing results from pStream 300 MB show extremely close report to pQRNG for all testing cases. It is hard to say wich testing data set is more random within the acceptable p-value between 0.01 and 0.99 of Chi Square. However, serial correlation value is worth to a close look because it indicates the correlation between each byte and its previous byte. pStream's serial correlation is $-4.0 \times 10^{-5}$ for 300 MB and $1.7 \times 10^{-5}$ for 1 GB, but pQRNG's serial correlaion is $8 \times 10^{-6}$. That means, pQRNG demonstrate slightly less serial correlation than qStream QRNG in this comparison.

Table 4. ENT randomness testing is tabulated for comparisons between physical QRNG from Quintessece Labs' pStream and pQRNG. We list testing reports from two data sizes from pStream.

| ENT | pStream 300MB | pStream 1GB | pQRNG 1GB |
|---|---|---|---|
| Entropy (bits) | 8.000000 | 8.000000 | 8.000000 |
| Chi Square | 231.03 | 259.41 | 231.04 |
| p-Value | 0.86 | 0.41 | 0.86 |
| Arith. Mean | 127.5035 | 127.5016 | 127.4995 |
| Monte Carlo $\pi$ | 3.14141912 | 3.14147598 | 3.141659557 |
| Serial Correlation | -0.00004 | 0.000017 | 0.000008 |

Figure 2 plots a slight variation of a deterministic pQRNG shown in Figure 1, used to create a quantum entropy booster or qeBooster. As an entropy booster, qeBooster injects the entropy to improve input prng's randomness. Linux /dev/random is an HRNG taking entropy from the system hardware. A typical Linux /dev/random has an entropy pool of 4096 bits. If the pool is not full, any random number request would be blocked utill the pool is full. In order to allow non-blocking random number generation, /dev/urandom is created. Based on that, urandom is a special PRNG associated with /dev/random. If the entropy pool is always full, then urandom would demonstrate excellent randomness, but the situation would be extremely bad if the pool is always nearly empty, which may be the case of cloud servers. In the case of servers are extremely lacking of entropy, they would generate keys with low entropy so reduce security. In this case, /dev/urandom can be piped with qeBooster to boost its entropy.

Other PRNGs can be also piped with qeBooster to boost its entropy for cryptographic pseudo random number generation. As an example, we take a popular fast PRNG created from C library rand() as the input prng for qeBooster.

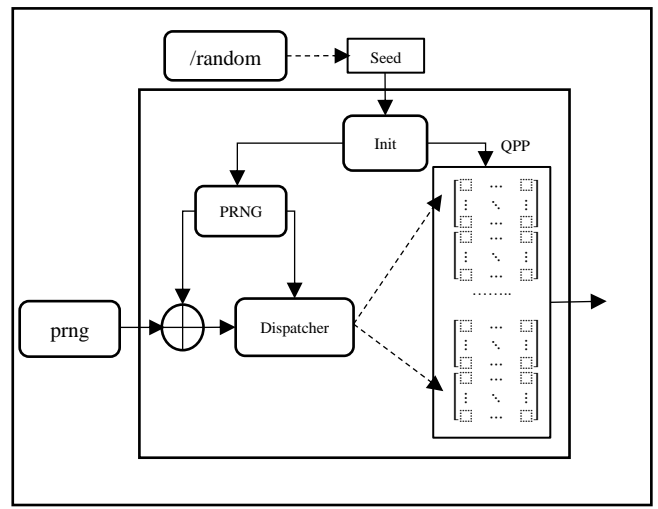

Figure 2. qeBooster behaves as an entropy booster for a low entropy pseudo random number generator.

Table 5. NIST testing is tabulated for rand() with qeBooster as its entropy booster.

| NIST 800-22 | Rand() | with qeBooster |
|---|---|---|
| Frequency | Success | Success |
| Block Frequency | Success | Success |
| Cumulative Sums | Success | Success |
| Runs | Success | Success |
| Longest Run | Success | Success |
| Rank | Success | Success |
| FFT | Failed | Success |
| Non-Overlapping Template | Success | Success |
| Overlapping Template | Failed | Failed |
| Universal | Success | Success |
| Approximate Entropy | Success | Success |
| Random Excursions | Success | Success |
| Random Excursions Variant | Success | Success |
| Serial | Success | Success |
| Linear Complexity | Success | Success |

Tables 5 - 7 demonstrate the testing results with rand() and rand() + qeBooster. After qeBooster, the boosted pseudo random numbers from rand() demonstrate good randomness improvements:

- one failure cases are disappeared in NIST testing;
- two failure cases in Dieharder are also disappeared, plus one weak case is reduced;
- in ENT tests, the major improvement is the Chi square from 190.91 with p-value = 1.00 to 284.43 with p-value = 0.10; the byte level bias is significantly improved. (note: in comparison wih Table 3, we notice that pseudo random numbers generated by rand() at different time show different randomness at different time.)
- in Dieharder test, rand() shows 6 weak and 2 failed. But after qeBooster, 5 weak and 0 failed.

Table 4. Dieharder testing is tabulated for rand() with qeBooster as an entropy booster.

| Dieharder | rand() | Rand() + qeBooster |
|---|---|---|
| Passed | 106/114 | 109/114 |
| Weak | 6/114 | 5/114 |
| Failed | 2/114 | 0 |

Table 5. ENT testing is tabulated for rand() with qeBooster as an entropy booster.

| ENT | rand() | Rand() + qeBooster |
|---|---|---|
| Entropy (bits) | 7.999999 | 7.999998 |
| Chi Square | 190.91 | 284.43 |
| p-Value | 1.00 | 0.10 |
| Arith. Mean | 127.4939 | 127.5054 |
| Monte Carlo $\pi$ | 3.141834126 | 3.141980526 |
| Serial Corr. | 0.000007 | 0.000022 |

qeBooster comes with huge entropy, over 100Kb, and adapts QPP as its entropy injection algorithm. It can be applied to any input data, even with statistically biased plaintexts. We want to demonstrate its capability with a 100MB of English characters to see how powerful it would be to blend any data into randomness. Table 8 tabulates its testing results with ENT. The plaintext file fails all ENT test cases:

- The entropy per 8-bits is 4.22, indicating the input data are indepent English sympols [25].
- Chi square is 1821992676.77 with p-value 0.0001, meaning totally bias.
- Arithematic mean is 97.9686, but ideal value is 127.5.
- Monte Carlo $\pi$ value is 4.00 not 3.14159265, so a unit square not a unit circle.
- Serial correlation is -0.138722, showing the strong correlation for each byte to its previous byte.

Then after qeBooster, the output file demonstrates a good randomness for all ENT test cases:

- The entropy per 8-bits is 7.999998, no longer English characters, with 0% compression rate.
- Chi square is 233.20 with p-value 0.83, no visible byte level bias existed.
- Arithematic mean is 127.4953, very close to ideal 127.5.
- Monte Carlo $\pi$ value is 3.141981640, with error ~0.01%.
- Serial correlation is $-9.3 \times 10^{-5}$, dropped down from -0.139.

It is clearly seen from this extreme case that qeBooster injects great entropies into input data and make it be in good randomness, thanks to quantum permutation pad. In this case, qeBooster acts as a data encryptor with the boosted data as the ciphertexts of input plaintexts. We also display the testing results with qStream QRNG with 200 MB random data as our comparison to the output of qeBooster. Both sets of data show close randomness to each other.

Table 8. ENT testing is tabulated for statistically biased plaintext inputs with qeBooster as an entropy booster.

| ENT | Plaintexts | + qeBooster | qStream 200MB |
|---|---|---|---|
| Entropy (bits) | 4.224280 | 7.999998 | 8.000000 |
| Chi Square | 1821992676.77 | 233.20 | 240.45 |
| p-Value | 0.0001 | 0.83 | 0.73 |
| Arith. Mean | 97.9686 | 127.4953 | 127.501 |
| Monte Carlo $\pi$ | 4.000000000 | 3.141981640 | 3.14121543 |
| Serial Corr. | -0.138722 | - 0.000093 | -0.000004 |

qeBooster may be a good candidate for the whitening algorithm of QRNG or any HRNG. A physical quantum random number generator naturaly produce the output random numbers with certain biases. In order to remove the biases, a whitening algorithm must be used to produce true random numbers. John von Neumann invented an algorithm to discard all '00' and '11' bits and convert '10' to '1' and '01' to '0'. This algorithm works nicely but it directly wastes 75% of bits. It is possible to use qeBooster with extremely high entropy to "smooth out" the bias. Therefore, we would waste any valuable bits generated from a physical QRNG then the physical throughput of a QRNG can be increased by 4-8x.

## IV. CONCLUSION

This paper proposes to use quantum permutation pad or QPP as a fundamental building block for pseudo quantum random number generator or pQRNG, entropy booster for low entropy PRNGs and whtining algorithm for HRNGs including QRNGs to increase their physical random number generation speeds. pQRNG demonstrates excellent randomness in random number generations, with Giga bytes per second. As an entropy booster, it can dramatically improve the randomness of any input data. It has a small footprint at 2.5KB so it can be embedded in any system to boost system pseudo random number generations such as /dev/urandom in Linux. We will perform further benchmarking exploration in the near future.